\documentclass[12pt,preprint]{aastex}

\newcommand{\etal}{{\it et al. }}

\shorttitle{A New Method To Resolve X-ray halos}
\shortauthors{Yao et al.}

\begin{document}

\title{
A New Method to Resolve X-Ray Halos around Point Sources with Chandra Data 
and Its Application to Cygnus X-1}

\author{Yangsen Yao\altaffilmark{1,2}, Shuang Nan Zhang\altaffilmark{1,2}, Xiaoling Zhang\altaffilmark{1,2}, Yuxin Feng\altaffilmark{1,2}}

\altaffiltext{1}{Physics Department, University of Alabama in Huntsville,
Huntsville, AL 35899}
\altaffiltext{2}{National Space Science and Technology Center, 320 Sparkman Dr., SD50, Huntsville, AL 35805 \\
\hspace{0.5in} 
yaoy@email.uah.edu, zhangsn@email.uah.edu, zhangx@email.uah.edu, fengyx@jet.uah.edu}

\begin{abstract}
With excellent angular resolution, good energy resolution and broad energy
band, the Chandra ACIS is the best instrument for studying the X-ray halos
around some galactic X-ray point sources caused by the dust scattering of
X-rays
in the interstellar medium. However, the direct images of bright sources
obtained with ACIS usually suffer from severe pile-up.
Making use of the fact that an isotropic image could be
reconstructed from its projection into any direction, we can reconstruct
the images of the X-ray halos from the data obtained with the 
HETGS and/or in CC mode.
These data have no or less serious pile-up and enable us to
take full advantage
of the excellent angular resolution of Chandra. With the reconstructed
high resolution images, we can probe the X-ray halos
as close as 1$''$ to their associated
point sources. Applying this method to Cygnus X-1 observed with Chandra
HETGS in CC mode, we derived an energy dependent radial halo flux distribution
and concluded that, in a circular region (2$'$ in radius) centered at 
the point source:
(1) relative to the total intensity, the
fractional halo intensity (FHI) is about 15\% at
$\sim$1~keV and drops to about 5\% at $\sim$6~keV; (2) about 50\%
of the halo photons are within the region of a radius less than 40$''$; and (3) the spectrum of
the point source is slightly distorted by the halo contamination.
\end{abstract}

\section{Introduction}
Small-angle scatterings between X-rays and dust grains
 in the interstellar medium (ISM) form halos, the diffuse emission around
X-ray point sources. The spectrum and the intensity distribution of
an X-ray halo
depend on the properties of the ISM along the line-of-sight,
i.e., the density distribution of the dust,
the grain size of the dust and the chemical composition of the grains
(Overbeck 1965; Hayakawa 1970; Martin \& Sciama 1970; Mathis, Rumpl \&
Nordsieck 1977; Predehl \& Klose 1996).
Because of the ISM absorption and scattering,
the X-ray energy band provides advantages over other
wave bands in studying the interstellar dusts.

The existence of the dust scattering X-ray halos was first discussed by
Overbeck (1965) and was first observationally confirmed by Rolf (1983)
using the data of GX339-4 with 
IPC/Einstein.
Using HRI/Einstein,
Catura (1983) and Bode \etal (1985) also confirmed the existence of
X-ray scattering
halos around point sources GX 3+1, GX 9+1, GX 13+1, GX 17+2 and Cygnus X-1.
The recent main results on X-ray halo studies were reported by
Predehl \& Schmitt (1995) by
systematically examining 25 point sources and 4 supernova remnants with 
ROSAT observations.

In all these investigations, the analyses were performed by
subtracting the point source surface brightness predicted by the point
source flux and the point spread function
(PSF) of the instrument from the observed
surface brightness, then comparing the inferred fractional halo intensity 
(FHI) as
a function of the off-axis angle to those predicted by
different halo models in order to deduce the halo properties.
However, because the differences
between halo
profiles from different spatial distributions and dust models
are significant only in the core of the halo (Mathis \& Lee 1991),
the limited angular resolutions of previous instruments
(1$'$ for IPC/Einstein and 25$''$ for PSPC/ROSAT)
prevent the previous works from probing regions 
close to the point sources, and make it difficult 
to study the properties of the dust grains and
to distinguish
between various dust models observationally
(Predehl \& Klose 1996).

The Advanced CCD Imaging
Spectrometer (ACIS) aboard the Chandra X-ray Observatory,
with its excellent angular resolution (0.5$''$ FWHM in PSF),
broad energy band ($0.2-10.0$~keV) and reasonably good energy
resolution ($E/\Delta E=10-60$), is the most
promising instrument to date in the X-ray halo study.
However, the timed exposure (TE) mode of ACIS
\footnote{For the Chandra instruments (ACIS, HETG etc.) 
and observation mode (CC mode, TE mode etc.), please refer to
http://cxc.harvard.edu/proposer/POG.}, which is the mode
with two-dimensional (2-D) image, often suffers from severe pile-up
because of the long exposure time (0.2 to 10 seconds) per frame
(pile-up is caused by two or more photons impacting one pixel or several
adjacent pixels in a single frame);
a bright X-ray point source will ``burn''
a hole at the source position
because most photons from the source are lost.
It is difficult to estimate the source flux and
the halo profile near the point source (see, e.g., Smith, Edgar
\& Shafer 2002) in this case.
Because only bright sources can generate significant
X-ray scattering halos, and
observations of bright sources with ACIS/Chandra TE mode always
suffer from severe pile-up,
the excellent spatial resolution of Chandra
has not brought the anticipated breakthrough
to X-ray halo study.

The pile-up can be avoided or lessened via reducing the number of 
photons impacting one pixel or several adjacent pixels 
in a single frame time. 
The Continuous Clocking (CC) mode
uses a very short frame time, and the transmission grating spreads
photons in different energies to different pixels; they can provide
us with pile-up free or less piled-up data.
However, the CC mode data only provide one-dimensional (1-D)
intensity distribution. 
The grating data, though provide
2-D information, are already dispersed by the
grating instruments.
In this letter,
we propose a new method to reconstruct the X-ray scattering halos
associated with X-ray point sources from
the CC mode data and/or the transmission grating data.
This method enables us to probe
the halo intensity distribution in a broad energy band
as close as 1$''$ to the point source. After testing it
with the MARX simulation, we applied
this method to Cygnus X-1 observed with 
the Chandra High Energy Transmission
Grating Spectrometer (HETGS, or ACIS with HETG) in CC mode.

\section{METHOD AND SIMULATION}
\begin{figure}
\plotone{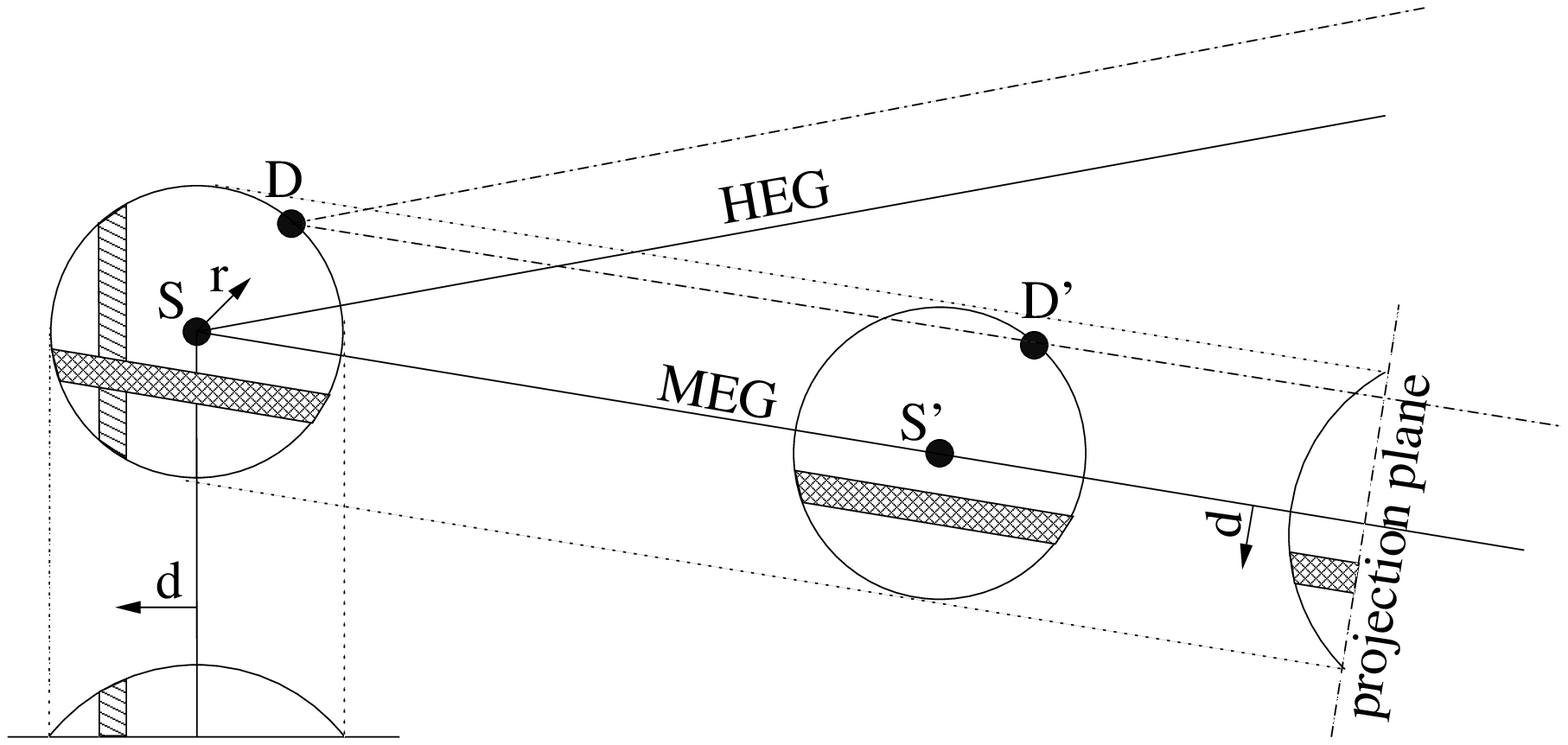}
\caption{
The projection of the photons along the grating arm and the projection 
of the photons in
zeroth order image along an arbitrary direction.
}
\label{method}
\end{figure}

After reflected by the Chandra mirror, X-rays
will be diffracted by the transmission grating (in one dimension)
by an angle $\beta$ according to the
grating equation,
\begin{equation}
p \sin\beta = m \lambda,
\end{equation}
where $p$ is the spatial period of the grating lines, $\beta$ is the
dispersion angle, $m$ is the grating order
($0, \pm1, \pm2, \cdots$),
and $\lambda$ is the photon wavelength.
The zeroth order image ($m=0$) is the same as the direct image
except for a smaller flux, because some photons are diffracted 
to higher orders.
For a mono-energy source, Chandra
transmission grating will detect exactly the same 
intensity
distribution in its higher order images
as in its zeroth order image, as
long as the source size is not too large
(less than 3$'$), as shown in Fig.~\ref{method}.
If we project the secondary and higher order photons
to a line perpendicular to the grating arm, the projected 1-D image
will be the same as the projection of the zeroth order image, except for
some broadening caused by ``mis-aligned'' grating facets of the HETG
in the cross-dispersion
direction
(see following discussion).
Because the grating only diffracts photons along
the direction of the grating arm, the above projection is also
valid for sources with continuum spectra.
Usually the non-zeroth order grating images and the CC mode data 
have no or much less pile-up;
either of them can be used to reconstruct the original image.

If the flux of a point source plus its X-ray halo
is isotropically distributed and centered at the point source as
$F(r)$,
and the projection process described above
can be represented by a matrix operator $M(r, d)$,
then the projected flux distribution $P(d)$ is 
\begin{equation}
P(d) = F(r)\times M(r, d),
\end{equation}
where $r$ is the distance from the centroid source position and
$d$ is the distance
from the projection center (refer to Fig.~\ref{method}).
The inverse matrix of the operator $M(r,d)$ exists, 
and the original distribution can
be easily resolved.
We used numerical integration to approach the above
projection process and built a matrix to approximate
the integration and calculate $M(r,d)^{-1}$.

\begin{figure}
\plotone{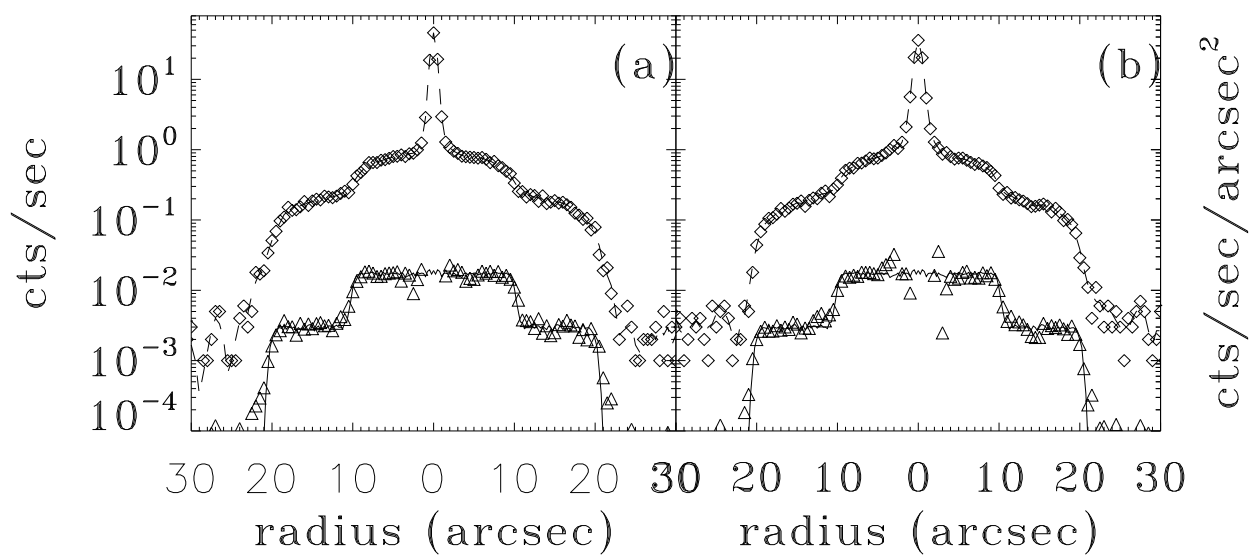}
\caption{
The reconstruction of the intensity distribution of
a simulated X-ray point source with halo
in the energy band 1.0--1.5~keV. 
Panel~(a): zeroth order;
panel~(b): MEG negative orders.
The curve with diamond symbols 
and the dashed line
are the projected photon distribution (cts/sec).
The triangle symbols are for the reconstructed halo
distribution (cts/sec/arcsec$^2$), 
after subtracting the PSF of the point source.
The solid line is the halo distribution from the zeroth order image
of the simulated halo (no pile-up in the simulation).
}
\label{simulation}
\end{figure}

To test our method, we produced with MARX 3.0
simulator\footnote{http://space.mit.edu/ASC/MARX} 
a point source plus two disk sources to mimic a point source
with its X-ray halo observed with Chandra/HETGS
in TE mode. Using the intrinsic CCD energy
resolution, we extracted the photons in the energy range 1.0--1.5~keV
and performed the test in this energy band.
We projected the zeroth order photons 
along an arbitrary direction to mimic the CC mode,
projected the MEG photons along the MEG grating arm,
and then multiplied these projected flux distribution with $M(r,d)^{-1}$
to resolve the flux distributions of the point source plus its halo. 
The PSF for the grating data was obtained with the same procedure,
from a simulated point source. It is worth noting that MARX
simulator takes into account the alignment blurs due to ``mis-aligned''
grating facets, so the broadening effects 
mentioned before will not affect the reconstruction results. 
For the CC mode data, the simulated point source is used directly as the PSF.
The halo flux 
was obtained by subtracting the corresponding PSFs from the flux of source plus
halo. The reconstructed halo surface brightness distribution is 
consistent with the simulation input, as shown in Fig.~\ref{simulation},
so is the FHI, with the input value of 27.7\% and 
the recovered value of 25.5\%.
We therefore conclude that the proposed method is
feasible to resolve the intensity distribution
of X-ray halos associated with point sources.


\section{APPLICATION TO CYGNUS X-1}
Cygnus X-1, the first dynamically determined X-ray binary system
to harbor a black hole,
had been observed seven times with Chandra by 2003 February 14.
It is so bright that during each of the total four observations with TE mode,
there was significant piled-up not only in the zeroth order image
but also in the dispersed grating images. The only short-frame observation
(1999 October 19, ObsID 107) was almost pile-up free in the grating image,
but the statistical quality of the data was poor.
The other three observations used CC mode. 
We applied our method to the one
with the highest statistical quality, which was observed on 2000
January 12 (ObsID 1511) with effective exposure of 12.7 ks.

We used the zeroth order data within 2$'$ from the source position.
For energies above 3.0~keV, the grating arms extend into the
2$'$ region; therefore we must exclude the photons resolved as grating
(non-zeroth order) events, which inevitably include some mis-identified 
halo photons and leave a gap in the halo intensity distribution. 
We interpolated across the gaps with an exponential function.
To estimated the pile-up in the zeroth
order data, we calculated the source flux from the grating arm and input it to
PIMMS\footnote{http://cxc.harvard.edu/toolkit/pimms.jsp} and obtained about
23\% pile-up. 
To estimate the point source flux more accurately, we compared
the grating data of the short-frame observation on 1999 October 19 with
the grating data of this CC mode observation, then used the read-out streak
of the short-frame to infer the point source flux at different energy bands
(with width 0.5~keV). 
We also input the grating spectrum of this observation
to MARX (scaling up the frame time by three orders of magnitude 
and scaling down the source flux accordingly to mimic the CC mode) 
to obtain the piled-up PSFs for different energy bands.

\begin{figure}
\plotone{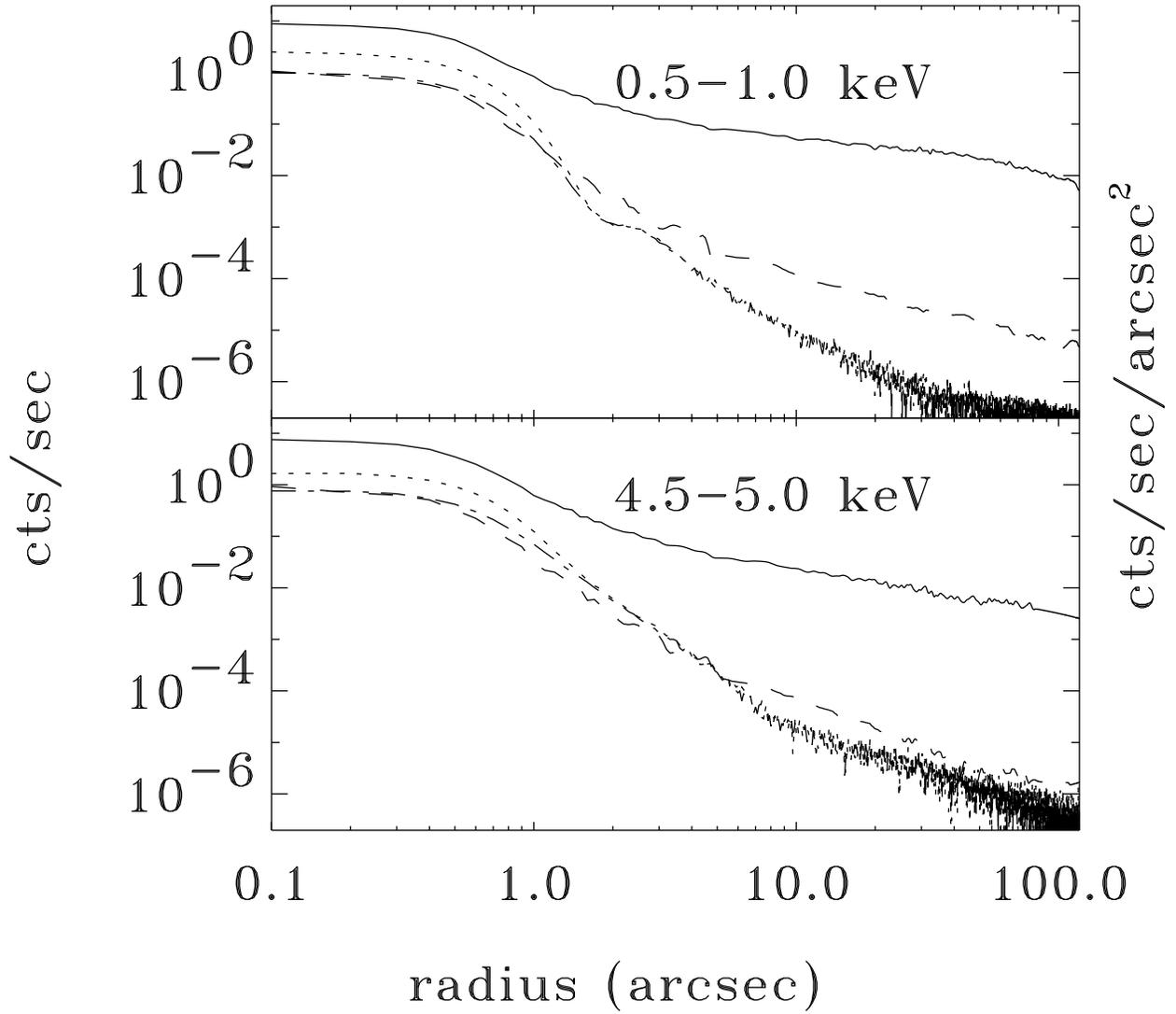}
\caption{
The flux distribution of Cygnus X-1 
and the PSF in two energy bands. In each panel, the
solid line indicates the projected flux distribution (cts/sec),
and the other lines are for flux distributions (cts/sec/arcsec$^2$).
Dashed line: source with halo; 
dotted line: PSF;
dash-dotted line: piled-up PSF.
}
\label{cygx1}
\end{figure}

The reconstructed flux distributions in two energy bands
are shown in Fig.~\ref{cygx1}. Even though there is pile-up 
in the core region,
the halo can be clearly resolved
down to 1$''$ from the point source in the 0.5--1.0~keV band.
We also fitted the reconstructed image with the simulated piled-up PSF
and then estimated the halo flux.
The results of
these two methods agree with each other.
The total FHI within 2$'$ as a function of photon
energy is shown in Fig.~\ref{ratio}(a); the fraction drops from about 15\% 
around 1~keV to about 5\% around 6~keV.
We define the half-flux radius $R_{0.5}$
of the halo as the radius which encloses half of the halo photons 
in the 2$'$ region.
The half-flux radius $R_{0.5}$ as a function of energy is 
shown in Fig.~\ref{ratio}(b);
clearly 50\% of the halo photons
are concentrated within 40$''$.
We also investigated how the halo contaminated the point source spectrum
in Cygnus X-1 
(see Fig.~\ref{ratio}(c) and Fig.~\ref{ratio}(d)). 
The halo spectrum
is softer than the point source spectrum, especially in the low energy band
(below 3~keV). Contributing only $\sim$ 10\% to the total brightness,
the X-ray halo in Cygnus X-1 does not distort the original
point source spectrum significantly.


The accurate Chandra PSF is important in evaluating FHI.
We use MARX simulator instead of the CIAO tool $mkpsf$
to generate the PSFs, because the PSF library grids,
from which $mkpsf$ interpolates to get a PSF for 
a certain off-axis angle and
a certain energy, are very coarse.
More realistic PSFs\footnote{http://cxc.harvard.edu/chart/index.html}
could be generated by producing PSFs of the mirror with 
the SAOSAC ray-tracing code
and then projecting
the simulated rays onto the detector in MARX. The MARX HRMA model
is a simplified version of SAOSAC and 
is proved sufficient for this study.

We used two previous Chandra observations 
to make consistency check of the PSF generated by the MARX simulator;
one is the the calibration observation of Her X-1 on 2002 July 1, 
the other is the long-frame Chandra/HETGS
observation of Cygnus X-1 on 1999 October 19.
Her X-1 is a ``real'' point source and almost halo free. 
We did a quick analysis of the Her X-1 data and found that beyond
15$''$ pile-up is negligible. The energy dependent
radial flux of Her X-1 is used it as the PSFs at off-axis
 angle 15$''$ to 120$''$; it gave the same results as the simulated PSFs.
For the core region of the PSF, we projected the read-out streak of
the Cygnus X-1 data and reconstructed a radial profile at $\sim$arcsec region, 
which agrees with
the simulated PSF in the core region very well.

The same Cygnus X-1 data were used to double check the inferred FHI.
Assume the half-flux radius $R_{0.5}$ does not change between observations,
we calculated  the halo photon intensity from photons 
between the half-flux radius
$R_{0.5}$ and 2$'$. 
The source count rate was obtained 
from the read-out streak.
The derived FHI is
plotted in panel (a) of Fig.~\ref{ratio}. It agrees well with the result
given previously in this section (shown in the same plot).

\section{CONCLUSION AND DISCUSSION}

In this letter, we propose a new method
to reconstruct the image 
of an X-ray point source with its associated X-ray scattering halo
 from the CC mode data and/or grating data,
which is then used to resolve the X-ray halo.
With this method and the high angular resolution of the Chandra Observatory
we are able to probe the intensity distribution
of the X-ray halos as close as 1$''$ to their associated point sources.
This method is tested with the MARX simulation
and applied to Cygnus X-1.

The derived FHI is energy dependent,
but does not seem to follow the $E^{-2}$ law, which was observed by
Predehl \& Schmitt (1995) and Smith \etal (2002).
The discrepancy may be due to 
the limited region we used (within 2$'$), because of
the following two reasons: 
(1) the scattering optical depth
$\tau_{sca} \propto E^{-2}$ was derived by integrating over all solid angles
(Mathis \& Lee 1991), instead of the 2$'$ region we used; 
and (2) for
single-scattering the mean scattering angle is related to
energy as $E^{-1}$ (Mathis \& Lee 1991), which means we under-estimated 
low energy photons more than we did for high energy photons.
In the low energy band, the FHI we obtained
are reasonably consistent with
the value reported by Predehl \etal (1995) (11\% at the
ROSAT energy range 0.1--2.4~keV). 

\begin{figure}
\plotone{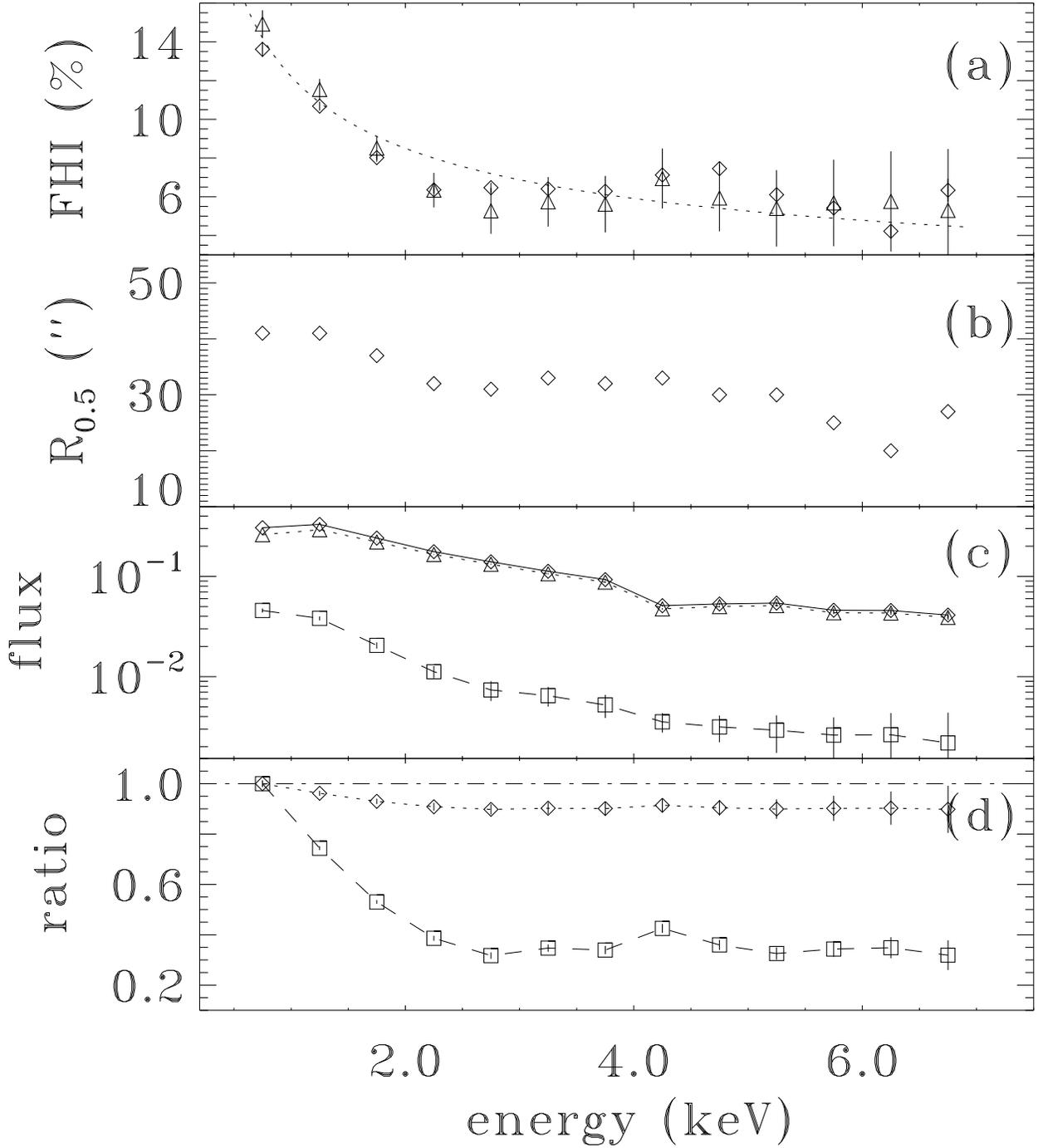}
\caption{
Panel~(a): Fractional halo intensity (FHI). 
Triangle symbols: result of our calculation;
diamond symbols: test result described in Section~3.
The dotted line indicates the best fit
$I(E) = (12.2\pm 0.6) (E/{\rm 1~keV}) ^{-0.52 \pm 0.05}$.
Panel~(b): the half-flux radius of the halo.
Panel~(c) spectra (cts/s/keV/$cm^2$), from top to bottom,
source with halo, the ``net'' point source, halo.
Panel~(d): the ratio of the normalized spectra.
Dotted line: source with halo to ``net'' point source;
dashed line: halo to ``net'' point source.
}
\label{ratio}
\end{figure}

The existence of the halo around a point source might distort the spectrum of
the point source.
Because the cross-section of the scattering process in the ISM is
energy dependent, the halo spectrum is different from the point source
spectrum (see Fig.~\ref{ratio}(c) and \ref{ratio}(d)).
Therefore if the instrument is unable to resolve
the point source from the halo, 
it will obtain contaminated source spectrum;
this is the case for X-ray instruments prior to Chandra.
Many of the previous measurements of the continuum X-ray spectra
of galactic X-ray sources with significant
X-ray scattering halo may suffer from this problem.
Despite that in Cygnus X-1 system, the X-ray scattering halo only contributes
about 10\% to the total brightness and does not distort the original spectrum
significantly, systematic studies of the X-ray halo distribution in
the broad band should be carried out for other
X-ray sources with significant X-ray scattering halos,
before we can draw any conclusion on the significance
of the halo induced distortion to their X-ray continuum spectra.

The spatial distribution of the dust grains is very important
in interpreting
the physical properties of the grains.
For single scattering, numerical
calculations show that
the core region of an X-ray halo is sensitive to the spatial
distribution of the scattering dust along the line-of-sight;
grains near the observer generate halos with flatter
profile near the core region, whereas grains close to the source create a peak
toward the central source; the wings, on the other hand, 
are not sensitive to the dust distributions
(Predehl \& Klose 1996; Mathis \& Lee 1991).
Because we can resolve the X-ray halos as close as 1$''$ to
the point sources, it is possible to determine 
the grain spatial distributions.
Even though the grain size distribution and
multiple-scatterings (especially for the low energy photons)
may reduce the differences between halo profiles from various
spatial distributions, this degeneracy can be easily resolved by drawing
the diagnostic diagram proposed by Mathis \& Lee (1991, Fig.~8).
The energy dependencies of the scattering optical depth
$\tau_{sca} \propto E^{-2}$ and the mean scattering angle
$\propto E^{-1}$, also make the broad band
energy-dependent FHI a good diagnostic of the
dust grains (Mathis \& Lee 1991).

In this letter,
we did not fit the energy dependent behavior of the halo with any
halo model to constrain the physical properties of the dust grains. However
with our technique, different grain spatial distributions and
different gain models can be well distinguished with further studies
of ACIS/Chandra data. We will address these issues in our next work.

\acknowledgments
Y. Yao thanks Allyn Tennant for useful discussions and insightful suggestions,
also thanks Daniel Dewey, Bruce Draine and the anonymous referee 
for helpful comments
and suggestions.
This work was supported in part by NASA Marshall Space
Flight Center under contract NCC8-200 and by NASA Long Term Space
Astrophysics Program under grants NAG5-7927 and NAG5-8523.

\clearpage

\clearpage

\clearpage

\clearpage

\end{document}